\documentclass[aps,pra, amsmath, showpacs, preprintnumbers,superscriptaddress,twocolumn,sort&compress,floatfix, amssymb]{revtex4}
\pdfoutput=1
\usepackage{graphicx}
\usepackage{rotating}
\usepackage{dcolumn}
\usepackage{bm}
\usepackage{color}
\usepackage{mathptmx, textcomp}
\usepackage[latin1]{inputenc}
\usepackage{braket}

\usepackage{multirow}

\begin{document}

\author{Michael Gr\"obner}\affiliation{Institut f\"ur Experimentalphysik und Zentrum f\"ur Quantenphysik, Universit\"at Innsbruck, 6020 Innsbruck, Austria}
\author{Philipp Weinmann}\affiliation{Institut f\"ur Quantenoptik und Quanteninformation (IQOQI), \"Osterreichische Akademie der Wissenschaften, 6020 Innsbruck, Austria}
\author{Emil Kirilov}\affiliation{Institut f\"ur Experimentalphysik und Zentrum f\"ur Quantenphysik, Universit\"at Innsbruck, 6020 Innsbruck, Austria}
\author{Hanns-Christoph N\"agerl}\affiliation{Institut f\"ur Experimentalphysik und Zentrum f\"ur Quantenphysik, Universit\"at Innsbruck, 6020 Innsbruck, Austria}

\title{Degenerate Raman sideband cooling of $^{39}$K}

\date{\today}


\begin{abstract}

We report on a first realization of sub-Doppler laser cooling of $^{39}$K atoms using degenerate three-dimensional Raman sideband cooling. We take advantage of the well-resolved excited hyperfine states on the D$_1$ optical transition to produce spin polarized samples with $1.4 \times 10^8$ atoms at temperatures of 1.8~$\mu$K. The phase-space densities are $\geq 10^{-4}$, which significantly improves the initial conditions for a subsequent evaporative cooling step. The presented cooling technique using the D$_1$ line can be adapted to other atomic species and is applicable to high-resolution imaging schemes in far off-resonant optical lattices.

\end{abstract}

\maketitle

\section{Introduction}
\label{Introduction}

The tremendous progress in laser cooling techniques has made the efficient production of Bose-Einstein condensates (BECs) for different atomic species feasible. Although evaporative cooling still serves as the final step to prepare degenerate atomic samples, efficient laser cooling prior to evaporation is indispensable. Starting from lower temperatures and higher densities after laser cooling has enabled the production of larger BECs within shorter preparation times, which has helped to shift the focus of ultracold atom experiments towards more ambitious goals such as probing of fundamental physics \cite{bloch2005,carr2009}, precision measurements \cite{ye2007}, quantum simulations of condensed-matter systems \cite{bloch2012}, and quantum computation \cite{Briegel2000,DeMille2002}.

In recent years, mixtures of quantum gases and the creation of polar ground-state molecules have attracted special attention. In particular, Bose-Fermi mixtures allow the study of e.g. novel supersolid phases \cite{Titvinidze08}, quantum phases that involve composite fermions \cite{Lewenstein04}, and mixtures under simultaneous superfluidity \cite{Ferrier-Barbut14}. Ultracold and dense samples of molecules produced out of Bose-Fermi mixtures \cite{Ni2008} promise access to new regimes of correlated quantum gases with novel quantum phase transitions, such as to topological superfluid phases~\cite{Cooper2009}. Due to the existence of fermionic and bosonic isotopes, the alkali-metal elements K and Li are prime choices for experiments on atom mixtures. Despite their popularity, laser cooling of these species remained challenging for a long time and has made the observation of, e.g., an optically trapped $^{39}$K BEC without an additional atomic coolant difficult \cite{Landini2012,Salomon2014,Groebner2016}. This is based on the fact that the excited hyperfine levels on the D$_2$ line are closely spaced, which complicates the implementation of sub-Doppler cooling schemes.

The first step forward towards lower laser cooling temperatures was the implementation of a multistage molasses cooling process in which the adiabatic sweeping of detunings and powers of two laser fields acting on the full ground-state hyperfine manifold allowed one to reach sub-Doppler temperatures $T \approx 25$~$\mu$K \cite{Landini2011,Gokhroo2011}. In a different approach, narrow-line laser cooling on a blue transition led to temperatures around 60~$\mu$K for K \cite{McKay2001} and Li \cite{Duarte2011}. Recently, gray molasses cooling (GMC) schemes, acting on the D$_1$ line, consisting of polarization gradient cooling and velocity-selective coherent population trapping (VSCPT) \cite{Aspect1988}, produced the so far coldest samples ($\approx 10$~$\mu$K) of K \cite{Fernandes2012,Salomon13,Nath2013} and Li \cite{Grier2013}. With this technique atoms are optically pumped into dark states, which couple depending on the atoms' velocity to bright states and reduce the scattering rate for atoms that are already cooled. Although the dark state suppresses the atom-light interaction for cold atoms, there is a weak coupling to bright states, which limits the final temperatures. Moreover, the scheme results in cold but unpolarized samples, requiring additional spin filtering or polarizing steps. Such measures typically come with an additional atom loss and an increase in the samples' temperatures.

Quite recently, the experimental challenge to implement single-lattice-site high-resolution imaging schemes in so-called quantum gas microscopes has attracted further attention to laser cooling schemes in the presence of optical lattices. Two competing schemes based on electromagnetically induced-transparency (EIT) cooling, which is similar to GMC in free space, and (nondegenerate) three-dimensional (3D) Raman sideband cooling have been proven to work for K \cite{Cheuk2015,haller2015} and Li \cite{Parsons2015}.

\begin{figure*}
	\begin{center}{
			\includegraphics[width=1.9\columnwidth]{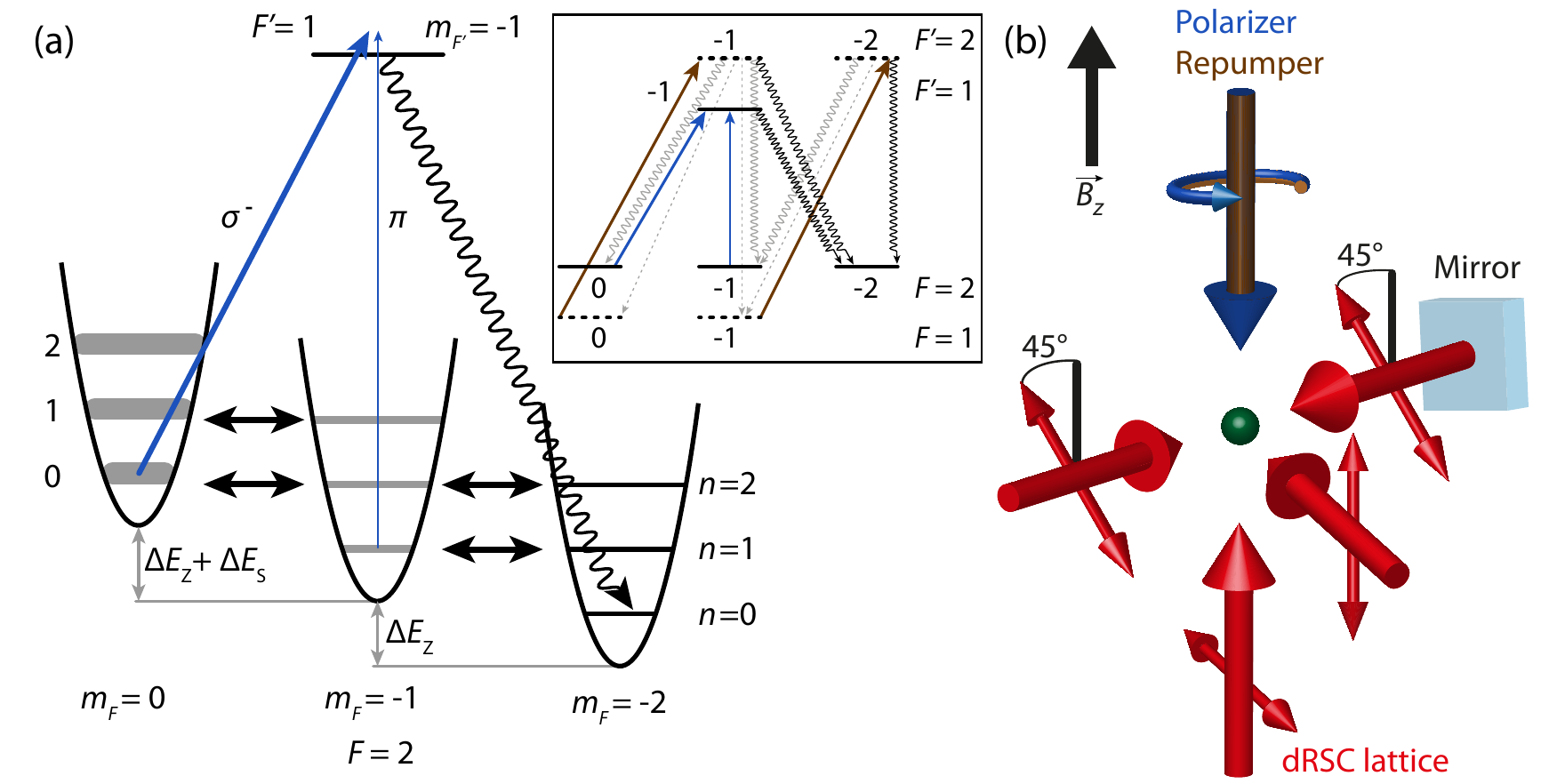}}
		\caption{\label{FIG1}(a) Schematic of dRSC on the D$_1$ line on $^{39}$K. One cycle consists of two 2-photon Raman transitions (double-sided arrows) between vibrational and magnetic states that are brought to degeneracy by an external magnetic field ($\Delta E_{\mathrm{Z}}$) and optical pumping (single-sided blue arrow) via the $\ket{F'=1, m_{F'}=-1}$ state. The strong $\sigma ^-$ component of the pumping beam broadens and shifts ($\Delta E_{\mathrm{S}}$) the energy levels. The inset shows all the states and transitions involved in the cooling scheme. The arrows represent the repumper (brown), the spontaneous decay into states with $F=2$ (gray, wiggled), the spontaneous transitions to the lower ground state with $F=1$ (gray, dashed), and the decay into the dark state (black, wiggled). (b) Simplified optical setup. The lattice (red inward pointing arrows) is generated by one retro-reflected standing wave with 45° polarization with respect to the magnetic field axis $\vec{B}_z$ and two running waves with polarizations that all lie in one plane. The repumper and polarizer beams (blue-brown arrow) propagate parallel to $\vec{B}_z$ to allow for a dominantly circular polarization.}
	\end{center}
\end{figure*}

In this paper we report on an implementation of degenerate Raman sideband cooling (dRSC) of $^{39}$K exploiting the D$_1$ transition. The experimental scheme is based on the pioneering work on dRSC of Cs \cite{Han2000,Vuletic98,Kerman2000,Hamann1998}, which was instrumental in achieving BEC for Cs \cite{Weber2003,Kraemer2004}. By avoiding the narrow excited-state hyperfine splitting on the D$_2$ line, we combine the advantages of having a dark state and an optical lattice to suppress the limitations existing for bright and gray molasses. Our scheme is applicable to far off-resonant lattice configurations, simplifying the existing high-resolution imaging techniques. Owing to the reduced scattering rate compared to GMC, we produce clouds with temperatures as low as 1.3~$\mu$K. For our largest atom numbers ($1.4 \times 10^8$) we obtain peak phase-space densities (PSDs) of spin-polarized samples of $\geq 10^{-4}$, which greatly improves the starting conditions for evaporative cooling. Furthermore, the demonstrated scheme is favorable for experiments with dual species involving K and some other species, e.g., Cs \cite{Groebner2016}, allowing a parallel production sequence due to the similarity in PSDs, and single-state preparation for both elements.

\section{Degenerate Raman sideband cooling}
\label{Ramancooling}

The dRSC scheme for $^{39}$K on the D$_1$ line is presented in Fig.~\ref{FIG1}(a). It has great similarities to the scheme that is widely used on the D$_2$ line of Cs \cite{Han2000,Vuletic98,Kerman2000,Hamann1998,Weber2003}. The main difference is a direct decay mechanism from the involved excited state to both hyperfine ground states, requiring an extra repumping beam. The atoms are harmonically confined at the wells of an optical lattice, which are depicted by parabolas. The figure shows the potentials for three different magnetic hyperfine sublevels $m_F$, which are energetically shifted with respect to each other by an external magnetic offset field to bring the $n$th vibrational sublevel to degeneracy with the $(n-1)$th vibrational sublevel of the next $m_F$ state (i.e., $\ket{F=2,m_F=-2,n=2}$ with $\ket{2,-1,1}$). The double-sided black arrows indicate degenerate 2-photon stimulated Raman transitions driven by the lattice beams to couple the degenerate states. The blue arrows represent a pumping beam on the $\ket{F=2}\rightarrow\ket{F'=1}$-D$_1$ transition (called the \textquotedblleft polarizer\textquotedblright), which carries a strong $\sigma^{-}$-polarized component and a very weak $\pi$ component. K atoms loaded into the lattice initially populate high-lying vibrational states $n$. During one cooling cycle the atoms are transferred from the $\ket{2,-2,n}$ to the $\ket{2,0,n-2}$ state by two 2-photon Raman processes that change the atoms' spin projection and vibrational state. The atoms get transferred by the $\sigma^-$ light into the $\ket{2,-2,n-2}$ state, completing one cooling cycle. In the Lamb-Dicke regime ($\eta=\sqrt{E_{\rm R}/\hbar\omega} \ll 1$; $E_{\rm R}$ is the photon recoil energy and $\hbar\omega$ the vibrational energy spacing) the spontaneous decay from the $\ket{F'=1,m_{F'}=-1}$ state conserves the vibrational state, which results in cooling by two vibrational quanta per cycle. The cooling continues until atoms reach either the $\ket{2,-2,0}$ or the $\ket{2,-1,0}$ state. The latter is cleared out by the weak $\pi$ component of the polarizer beam.

In contrast to the implementations of dRSC on samples of Cs and Rb on the respective D$_2$ lines, a transition from the excited ($\ket{F'=1,m_{F'}=-1}$) state to the second, lower-lying ground state ($F=1$) is possible. We find that a $\sigma^-$-polarized beam resonant with the $\ket{F=1}\rightarrow\ket{F'=2}$-D$_1$ transition leads to the lowest temperatures. With this polarization the number of states involved in the Raman scheme is minimal, keeping the steady-state scattering rates low. The inset in Fig.~\ref{FIG1}(a) illustrates the complete dRSC scheme as implemented in our experiment.

We use one retroreflected standing wave and two running waves with linear polarizations to generate the 3D lattice, similar to the original implementation on Cs. In that configuration, relative phase fluctuations between the separate lattice beams translate to rigid translations of the entire lattice, but do not affect the shape and the depth of the lattice sites and the polarization the atoms experience at each site \cite{Grynberg1993}. As shown in Fig.~\ref{FIG1}(b), the angles between the standing wave and the other two beams are close to \mbox{90°}. To maximize the Raman coupling, the polarizations of the running-wave beams are chosen to be linear and are made to lie in one plane. The polarizations of the standing wave are rotated by 45° with respect to the magnetic field axis $\vec{B}_z$ and also lie in the plane that is spanned by the polarizations of the running-wave beams. The polarizer beam and the repumping beam propagate parallel to $\vec{B}_z$ to allow for an almost perfect $\sigma^{-}$ polarization. The weak $\pi$ component is introduced by a slight tilt of the magnetic field axis.

\begin{figure}
	\begin{center}{
			\includegraphics[width=1\columnwidth]{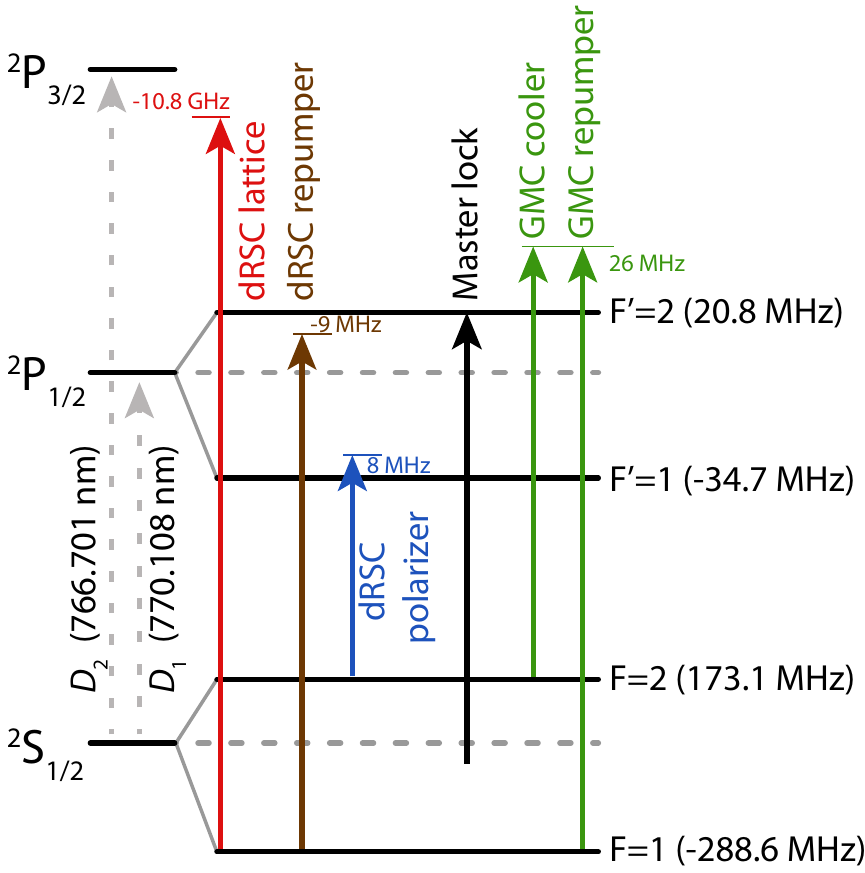}}
		\caption{\label{FIG2}Level scheme of the two lowest energy levels of $^{39}$K and transitions used for dRSC and GMC. The lattice laser is detuned from the D$_2$ line and the remaining beams for GMC and dRSC are derived from one laser locked on the D$_1$ line. The arrows show the transitions used and the numbers give the experimentally found optimum detunings.}
	\end{center}
\end{figure}

\begin{figure}
	\begin{center}{
			\includegraphics[width=1\columnwidth]{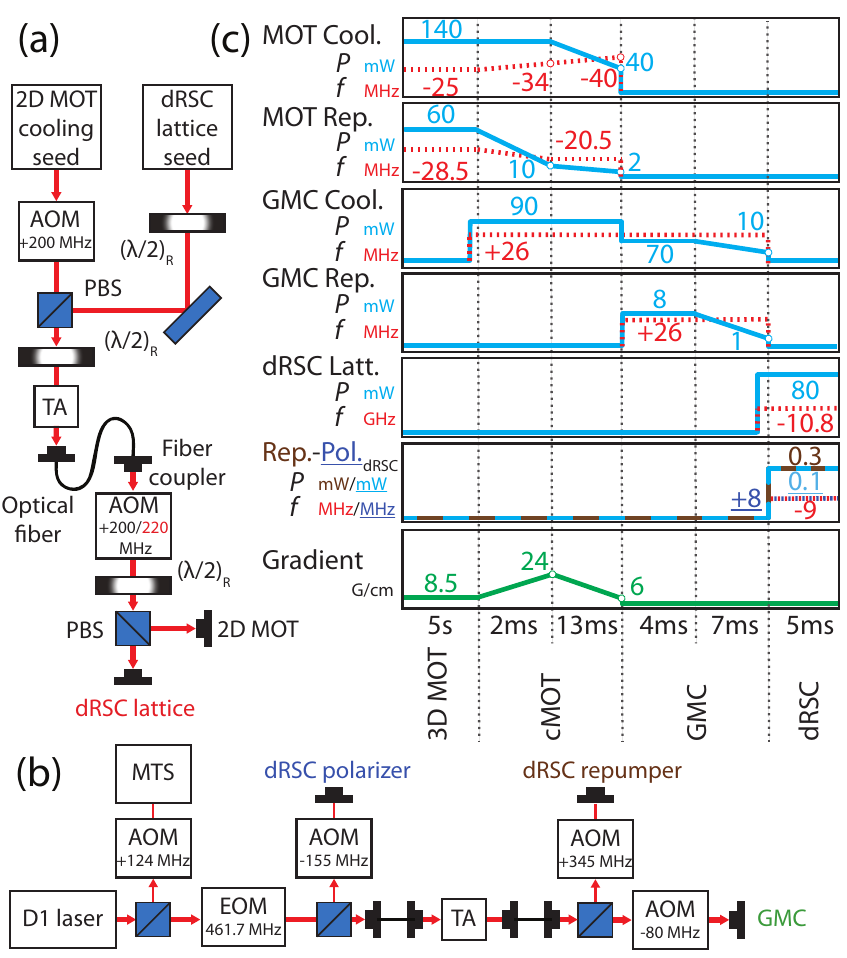}}
		\caption{\label{FIG3} (a) and (b) Optical setups to generate the light for dRSC. (c) Timing diagram for the $^{39}$K laser cooling sequence. The continuous (blue, green, brown) lines correspond to either total powers or magnetic field gradient strength and the dashed (red, dark blue) lines to frequency detunings (3D MOT: $\ket{F=2}\rightarrow\ket{F'=3}$-D$_2$ transition for cooling and $\ket{F=1}\rightarrow\ket{F'=2}$ transition for repumping).}
	\end{center}
\end{figure}

The lattice light is derived from a home-built interference-filter-stabilized external-cavity laser \cite{Baillard2006}. This laser design provides a broad tuning range. The separation of feedback and wavelength selection leads to a comparatively high-frequency stability, which avoids the need for active stabilization. The lattice light is several GHz detuned from the D$_2$ line (see Fig.~\ref{FIG2}) and amplified by a tapered amplifier (TA) gain chip. To save on infrastructure, this TA is the same one as the one that is used for generating the 2D-MOT cooling light. Home-built motorized $\lambda/2$ waveplates [denoted $(\lambda/2)_\mathrm{R}$ in Fig.~\ref{FIG3}(a)] dynamically change the power in the seed beams to the TA and allow for a redistribution of the light after the TA. Additionally, we take advantage of the already existing master laser that is needed for GMC on the D$_1$ line to provide the light for the repumper and polarizer beams. This laser is directly locked to the crossover transition of the two ground states to the $\ket{F'=2}$-D$_1$ state using modulation-transfer spectroscopy (MTS). In contrast to our implementation of GMC, where we imprint the repumping frequency on the carrier with a resonant electro-optical modulator (EOM), we shift the master laser's frequency with several acousto-optical modulators (AOMs) close to the resonance of the $\ket{F=1} \rightarrow \ket{F'=2}$-D$_1$ transition for the repumper beam and to the $\ket{F=2} \rightarrow \ket{F'=1}$-D$_1$ transition for the polarizer beam [see Fig.~\ref{FIG3}(b)]. This allows us to independently control the frequencies and powers of both beams.

We spatially overlap the repumper beam with the polarizer beam and send both beams via one optical fiber from the laser table to the main table, which carries the experimental chamber. With a beam diameter of 14.5 mm and powers of 0.4 and 0.2~mW, respectively, we achieve maximum intensities of 0.5 and 0.25~mW/cm$^{2}$ for the repumper and polarizer beams. The light for the lattice is sent via one fiber to the experimental table and there it is split into three parts, which all carry approximately one third of the available power. The lattice power is actively stabilized and we use a maximum total power of 85~mW (peak intensity of 705~mW/cm$^{2}$ per beam). The magnetic offset field needed to bring the different magnetic sublevels to degeneracy is produced by a pair of Helmholtz coils. Two independent cosine coils are used to compensate for stray fields along the horizontal directions and to slightly tilt the offset field with respect to the polarizer axis to control the strength of the $\pi$-polarized component of this beam.

\section{Experimental results}
\label{Experimentalresults}

The experimental setup is based on the apparatus presented in Ref.~\cite{Groebner2016}. In the experiments discussed here, we load up to $3 \times 10^8$ atoms in 5~s from a 2D MOT into a 3D MOT, which operates on the D$_2$ line. Differently from to our previous work, we skip the molasses cooling on the D$_2$ line and follow the cooling steps outlined in Ref.~\cite{Salomon13}, resulting in the sequence presented in Fig.~\ref{FIG3}(c). During the last 350~ms of the 3D MOT phase, we switch the seed light of one of our 2D MOT TAs to the Raman lattice light frequency to give the TA enough time to equilibrate before the actual Raman cooling takes place. Additionally, 50~ms before starting with the compressed MOT (cMOT), we quickly turn on the GMC light working on the D$_1$ line. As a next step, we simultaneously increase the magnetic field gradient of the MOT quadrupole field to 24~G/cm along the coil axis, reduce the power of the repumper beam to 10~mW, and increase the detuning of the MOT cooling beams to -34~MHz. After 2~ms, to increase the density of our sample and reduce the temperature, we ramp down the gradient to 6~G/cm in 13~ms, decrease the power of the repumper beam to 2~mW, detune the cooling light further to -40~MHz, and reduce the cooling power to 40~mW while having the D$_1$ line cooling light on during the whole procedure. With this technique we are able to reduce the temperature of our atom sample to 150~$\mu$K. We note that we cannot reproduce the densities reported in Ref.~\cite{Salomon13}. This might be connected to the fact that during the cMOT phase we cannot completely turn off the D$_2$ cooling light, since otherwise we observe a significant atom loss and a reduction of the atom density. To increase the PSD and reduce the requirements on our Raman lattice, we apply GMC on the D$_1$ line. We turn off the quadrupole field and imprint sidebands at the ground-state splitting of 461.7~MHz to tune the repumper and cooling lasers to the Raman resonance in a $\Lambda$ configuration. After 4~ms of constant power we linearly decrease the power of the beam to 10~mW within 7~ms. This process yields a cloud with a temperature of 8~$\mu$K. With no significant atom loss since the 3D MOT phase, we obtain a PSD of $1.2 \times 10^{-5}$.

We turn on the lattice light and the current that creates the offset field 1.5~ms before GMC ends to initialize dRSC. We estimate the initial offset field to be around $800$~mG. Next, we switch on the polarizer and repumper beams. Initially, for another 0.1~ms, we keep the offset field at a somewhat higher value than used later. This helps to compensate for the larger center-to-center spacing between the unbound energy bands above the lattice for nearly free atoms and ensures efficient cooling into the lattice potential \cite{Kerman2000}. During the cooling process we first linearly ramp down the magnetic offset field to $50$~mG in 0.5~ms and afterwards slowly increase it to $300$~mG to keep the magnetic sublevels degenerate. To characterize the efficiency of the dRSC we adiabatically release the atoms from the lattice and isolate the $\ket{F=2,m_F=-2}$ magnetic sublevel in a Stern-Gerlach experiment. This is done by applying a $7.5$-G/cm gradient and a 30-G offset field to levitate this state during expansion. After a hold time of 30~ms, the different spin components are well resolved and one can measure their temperatures individually. We find that some weak magnetic curvature along the vertical coil axis leads to weak trapping and hence prevents us from faithfully determining the temperature from the vertical expansion dynamics. We thus obtain the temperature from the expansion of the cloud in the horizontal directions. In these directions, the magnetic force that results from the levitation condition accelerates the atoms away from the gradient center. Hence all temperatures given here are upper bounds, but the deviation at a 30-G offset field is already comparatively small. In the following, we give cloud radii $\sigma_x$ at fixed expansion durations, which are proportional to the clouds' temperatures. The radii $\sigma_x$ are the standard deviations calculated by fitting Gaussian functions to the integrated optical densities. A cloud radius of $\sigma_x=1$~mm corresponds to a temperature of about 1.5~$\mu$K.

\begin{figure}
	\begin{center}{
			\includegraphics[width=1\columnwidth]{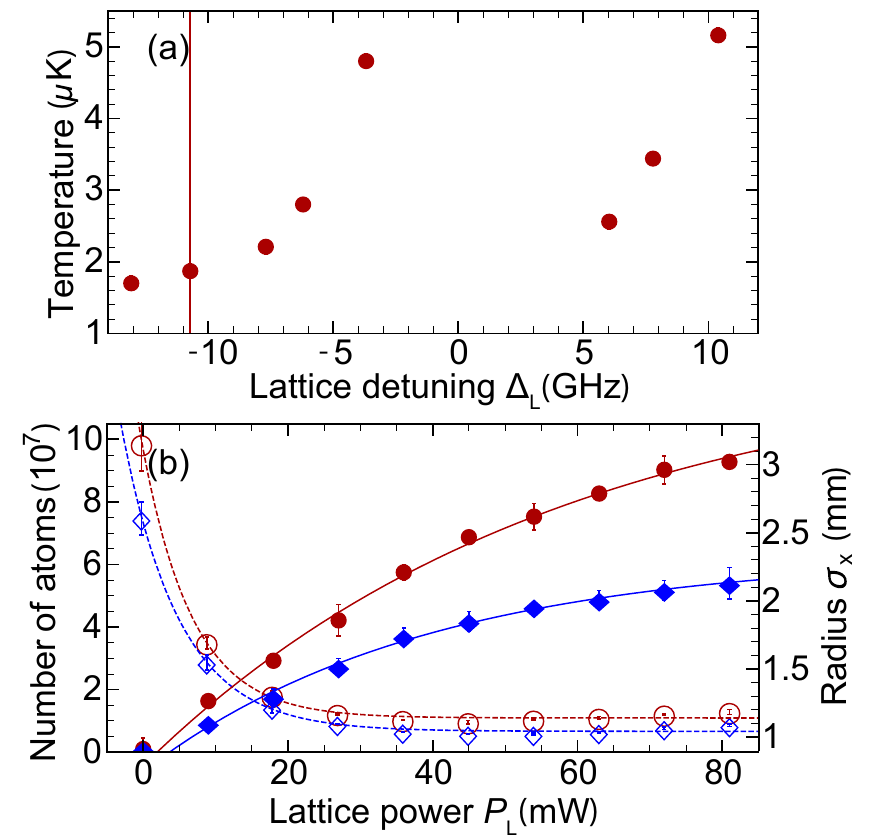}}
		\caption{\label{FIG4}(a) Temperature after dRSC as a function of the lattice detuning $\Delta_{\mathrm{L}}$ from the D$_2$ line at $P_\mathrm{L}=80$~mW. The vertical bar indicates the optimum detuning, which we find to be limited by lattice power. Error bars are smaller than the symbol size. (b) Number of atoms (solid symbols) and cloud radius $\sigma_x$ (open symbols) after dRSC as a function of the total lattice power $ P_{\mathrm{L}} $ for $1.2 \times 10^{8}$ atoms (blue diamonds) and $2.8 \times 10^{8}$ atoms (red circles) after GMC at $\Delta_\mathrm{L} = -10.8$~GHz. The solid and dashed lines are fits by a saturated growth function and an exponential decay function to guide the eye. The lowest temperatures are already reached at around 30~mW, whereas the capture efficiency still grows with increasing lattice power.}
	\end{center}
\end{figure}

We first discuss the influence of the lattice parameters on the dRSC cooling performance. For this, we keep the repumper and polarizer beams at constant detunings of 8 and -9~MHz, respectively. In general, the lattice light is responsible for trapping the atoms during the cooling process ($U_{\mathrm{dip}} \propto 1/\Delta_{\mathrm{L}}$) and providing the 2-photon stimulated Raman transitions that are needed to drive the cooling cycle. The lattice potential should be deep to operate dRSC in the Lamb-Dicke regime. In this regime spontaneous emissions as needed in the cycling process conserve the number of vibrational quanta. Also, the lattice light should cause minimal off-resonant excitations, which heat the sample ($\Gamma_{\mathrm{sc}} \propto 1/\Delta_{\mathrm{L}}^2$). Figure~\ref{FIG4}(a) shows the dependence of the temperature on the detuning $\Delta_\mathrm{L}$ from the D$_2$ line. For increasing blue detunings we observe higher final temperatures. In this regime the atoms that have already been cooled reside in potential minima where the light intensity is strongly reduced, leading to a suppression of off-resonant excitations, but also to a reduction of the Raman transition rates, which is the basis for further cooling. Overall, this appears to lead to a lower cooling efficiency. For red detunings, the final temperatures decrease with increased detuning. This is expected, because atoms are confined in intensity maxima, and for larger detunings, off-resonant excitations are more suppressed than Raman transitions. However, larger red detunings come with an enhanced particle loss, presumably as a result of the lattice depth. Since we are limited by our available total power, we find as a compromise an optimum detuning of $\Delta_\mathrm{L} = -10.8$~GHz, which is indicated in Fig. \ref{FIG4}(a) by a vertical line. At this detuning we obtain the highest PSDs after dRSC.

Next, we keep the detuning $\Delta_\mathrm{L}$ fixed and scan the lattice power $P_\mathrm{L}$ as we record the atom number and cloud radius after time of flight for two different initial atom numbers. The result is shown in Fig.~\ref{FIG4}(b). The dRSC process starts to work efficiently at powers larger than 30~mW, as can be seen from the greatly reduced cloud radii. Yet, the number of atoms has not saturated and we find that higher lattice powers than 30~mW are needed to cool the entire sample. With a peak intensity of 663~mW/cm$^2$ per beam, we calculate well depths of 60 and 30~$\mu$K, trap frequencies $\omega/(2\pi)$ of 159 and 72~kHz, Raman couplings $\Omega_\mathrm{R}/(2\pi)$ of 0 and 45~kHz, and Lamb-Dicke parameters of 0.23 and 0.35 along the standing-wave and free-running beam axes, respectively. The 45° polarization of the standing-wave beam with respect to $\vec{B}_z$ suppresses the Raman transitions along this axis but leads to an isotropic lattice potential along the two other axes. However, changing the polarizations from 45° breaks the condition of zero Raman coupling, but has no influence on the dRSC performance, as we find in the experiment. We believe that imperfections of our optical setup and the large Raman couplings in two dimensions, comparable to the average vibrational frequency, always break this condition and allow dRSC to work at one magnetic field value in all three spatial directions \cite{Kerman2000}.

\begin{figure}
	\begin{center}{
			\includegraphics[width=1\columnwidth]{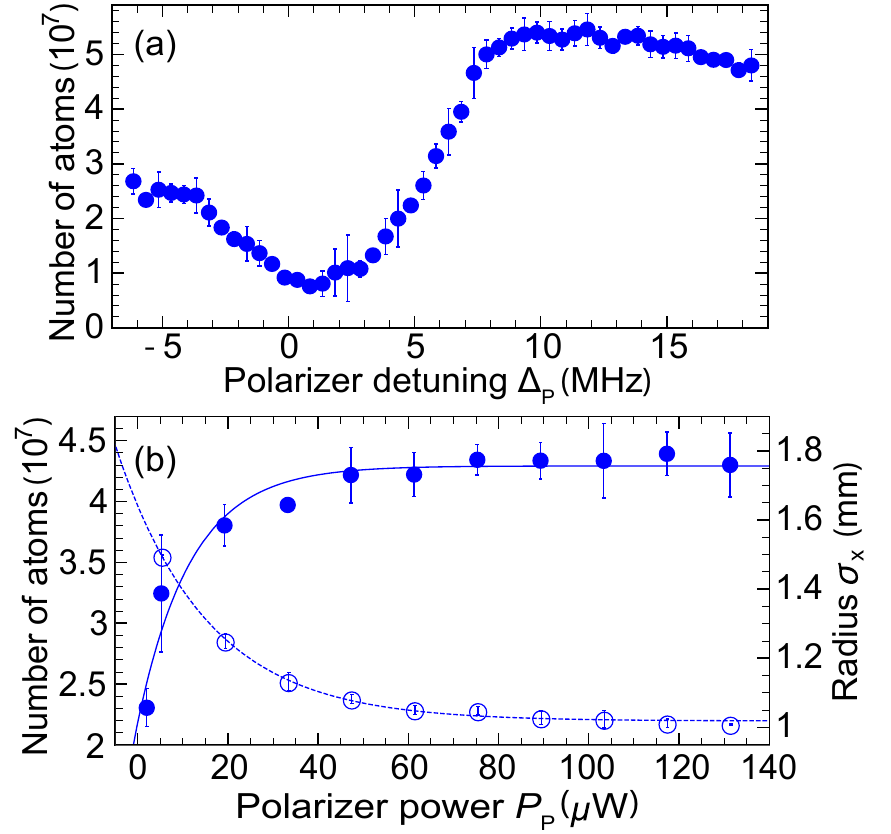}}
		\caption{\label{FIG5}(a) Number of atoms after dRSC as a function of the polarizer detuning $\Delta_\mathrm{P}$ at $P_\mathrm{P}=120$ $\mu$W. One can clearly see a minimum around zero detuning. (b) Number of atoms (solid circles) and cloud radius $\sigma_x$ (open circles) after dRSC as a function of the polarizer power $P_\mathrm{P}$ at $\Delta_\mathrm{P}=8$ MHz. The solid and dashed lines are fits by a saturated growth function and an exponential decay function to guide the eye. Already very small powers make dRSC work, but higher ones are needed to reach the lowest temperatures.}
	\end{center}
\end{figure}

We now study the influence of the polarizer's detuning $\Delta_\mathrm{P}$ and power $P_\mathrm{P}$ on the dRSC cooling performance. The results of our measurements are shown in Fig.~\ref{FIG5}. The number of atoms cooled and captured depends strongly on $\Delta_\mathrm{P}$. We observe a clear minimum for the atom number around zero detuning and a maximum when $\Delta_\mathrm{P}$ is blue detuned in the range from $+8$ to $+15$~MHz. We interpret our data in the following way: The strong $\sigma^-$ component causes a significant light shift $\Delta E_{\mathrm{S}}$ of the $m_F=0$ sublevel, as indicated in Fig.~\ref{FIG1}(a) for a blue detuning. The level gets shifted and also broadened by the near-resonant light. For blue detunings larger than $8$~MHz the effect of the light shift on the cooling process becomes small and negligible in view of a strong Raman coupling. Close to resonance the light shift breaks the degeneracy and hence limits the cooling performance. Towards larger red detunings the cooling process works again, however, not as well as on the blue side. We think that the opposite light shift increases the probability of driving Raman transitions that do not contribute to the cooling process.

Figure~\ref{FIG5}(b) shows that the power $P_\mathrm{P}$ needed to induce dRSC is very small (20~$\mu$W). To efficiently reach the final dark state and lowest temperatures, we need a significant portion of light with $\pi$ polarization ($\approx$5\% of $\sigma^-$) and therefore total powers of at least 60~$\mu$W. At $P_\mathrm{P}=120$~$\mu$W and $\Delta_\mathrm{P}=8$~MHz we expect a scattering rate $\Gamma_{\mathrm{S}}$ of 184~kHz for the polarizer. Another important property of the polarizer is that the free-running beam causes a net force that initially pushes the atoms out of the lattice. This could in principle be avoided by retroreflecting the beam at the cost of a less pure polarization and atoms staying in the nodes of the standing wave.

\begin{figure}
	\begin{center}{
			\includegraphics[width=1\columnwidth]{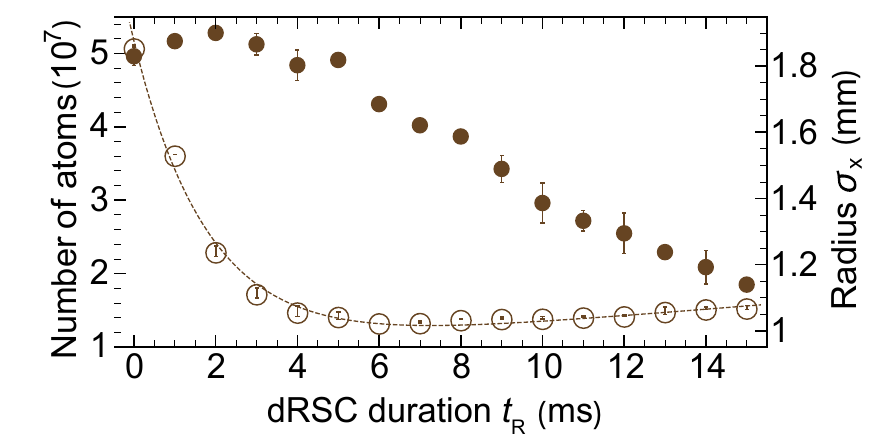}}
		\caption{\label{FIG6} Number of atoms (solid circles) and cloud radius $\sigma_x$ (open circles) after dRSC as a function of the cooling duration $t_\mathrm{R}$. The dashed line is a fit by an exponential decay function with a linearly rising offset to guide the eye. Cooling competes with atom loss, which results in an optimum cooling time of about 5~ms. At this point in time, the temperature is $1.6 \ \mu$K.}
	\end{center}
\end{figure}

In Fig.~\ref{FIG6} we study the dependence of the final temperature and atom number on dRSC duration $t_\mathrm{R}$. During the first 5~ms of cooling, the fraction of atoms in the $\ket{F=2,m_F=-2}$ state stays almost constant, whereas the temperature significantly decreases. For cooling times longer than 5~ms we observe a loss of atoms. Also, the radius slightly increases for longer cooling durations. To model the fast initial cooling we set up an effective four-level system, coupled by the lattice and polarizer light, and numerically solve the optical Bloch equations projected on it. Atoms initially in the $\ket{F=2,m_F=-2,n}$ state are pumped with a rate $\Gamma_\mathrm{cycle} \approx 3$~kHz into the $\ket{F=2,m_F=-2,n-2}$ state. In every cycle, an energy of $2\hbar\omega$ gets removed, resulting in a cooling rate $\Gamma_\mathrm{cool}$ of 24~$\mu$K/ms along the free-running lattice beam axes. Off-resonant excitations by the $\Delta_\mathrm{L}=-10.8$~GHz detuned lattice lead to a heating rate $\Gamma_\mathrm{heat}$ \cite{Grimm2000} on the order of 1~$\mu$K/ms. Although $\Gamma_\mathrm{cool} \approx 20  \Gamma_\mathrm{heat}$, we interpret the atom loss for $t_\mathrm{R}>5$~ms by the continuous interaction of the atom sample with the lattice light. Atoms populating higher lattice bands exhibit a larger probability to tunnel to other lattice sites, resulting in double occupancy or loss from the lattice volume by diffusion.

We finally comment on the importance of the repumping beam. While we see that some minimal intensity (more than 0.1 mW/cm$^2$) is needed to operate the cooling process, we find no strong dependence of the final temperature on neither the repumper detuning nor on its intensity. We find a slight minimum for the temperature for a red detuning of 9~MHz. All our data are thus taken at this value for the detuning.

Table \ref{table1} summarizes our results for two different starting conditions after the GMC stage. Similar to previous implementations of dRSC, we find some dependence of the final temperature on the initial density. With smaller atom samples ($N=5.6\times 10^{7}$ after dRSC) we realize temperatures around 1.6~$\mu$K (for even smaller samples with $N=2.5\times 10^{7}$ we have seen temperatures down to 1.3~$\mu$K) and for our largest samples ($N=1.4\times 10^{8}$) we measure $T \approx 1.8~\mu$K and obtain PSDs that are $\approx 1\times 10^{-4}$. In units of the recoil temperature $T_\mathrm{R}=\hbar^2 k_{\mathrm{L}}^2/m k_{\mathrm{B}}$ the lowest temperature that we observe is $T = 3.1$~$T_\mathrm{R}$. This is a factor of 2 larger than typical values for Cs and Rb \cite{kerman2002}. It is, to our knowledge, the lowest temperature achieved for K after laser cooling. Nevertheless, it would be of interest to understand the limits of dRSC on K. With a lattice detuning $\Delta_\mathrm{L}=-10.8$~GHz that is roughly comparable to what is used in the Cs and Rb experiments \cite{kerman2002}, other mechanisms than off-resonant excitations due to the lattice beams must be limiting the dRSC performance. More detailed studies are needed to quantify the relevance of the pumping into the other ground-state hyperfine level and the effect of reexcitation out of the dark state via the $\pi$ component of the polarizer in view of a smaller excited-state hyperfine splitting.

\begin{table}[]
	\begin{center}
		\caption{\label{table1} Comparison of the performance of dRSC for two different starting conditions after GMC (top and bottom). Fra: approximative percentage of atoms in the $\ket{F=2,m_F=-2}$ state; $N$: number of atoms in the $\ket{F=2,m_F=-2}$ state; $T$: temperature; $n_{\mathrm{p}}$: peak density; PSD: peak phase-space density in free space.}
		\begin{ruledtabular}
			\begin{tabular}{rccccc}
				$\space$                         & Fra (\%) & $N$               & $T \, (\mu \mathrm{K})$ & $n_{\mathrm{p}} \, (\mathrm{cm^{-3}})$ & PSD                  \\ \hline
				\noalign{\smallskip}
				GMC                          & --     & $1.3 \times 10^{8}$ & $8$                     & $1.5 \times 10^{10}$                                & $1.5 \times 10^{-5}$ \\
				dRSC                             & 83     & $5.6 \times 10^{7}$ & $1.6$                   & $5.9 \times 10^{9}$                                 & $6 \times 10^{-5}$   \\ \hline
				\noalign{\smallskip}
				GMC                          & --     & $3 \times 10^{8}$   & $8$                     & $1.3 \times 10^{10}$                                & $1.2 \times 10^{-5}$ \\
				dRSC                             & 78     & $1.4 \times 10^{8}$ & $1.8$                   & $1.1 \times 10^{10}$                                & $1 \times 10^{-4}$   \\
			\end{tabular}
		\end{ruledtabular}
	\end{center}
\end{table}

We have additionally investigated variations of dRSC that reduce the required optical infrastructure. For example, we have imprinted the polarizer on the repumping beam using the resonant EOM that is also used for GMC. We have found minimum temperatures around 2~$\mu$K. We attribute such higher temperatures to the reduced flexibility in choosing the individual frequencies and intensities. Also, repumping on the D$_2$ line via the 3D MOT beams has resulted in higher final temperatures around 3.5~$\mu$K. Here, less control over the polarization complicates the repumping process. Finally, substituting D$_1$ by D$_2$ molasses cooling has led to a larger atom loss, but it has not influenced the final temperatures.

\section{Conclusion}
\label{Conclusion}

We have experimentally demonstrated that dRSC on the D$_1$ line can be used to efficiently produce ultracold samples of $^{39}$K. With the scheme presented in this paper, we have measured a fourfold reduction in temperature and a tenfold increase in PSD compared to our results for GMC, resulting in the lowest temperatures observed so far for K and the highest PSD for a spin polarized K sample after laser cooling. Moreover, the high degree of polarization avoids the necessity of spin polarization and the high PSD connected with the low temperature should allow for direct loading of a large volume dipole trap, similarly to schemes successfully realized for Cs \cite{Weber2003} and Rb \cite{Pilch2009}. In that trap, radio-frequency adiabatic rapid passages can transfer the atom sample into the absolute ground state. Finally, in contrast to our previous strategy \cite{Groebner2016}, dRSC opens the route to simultaneously cool K in the presence of, e.g., Cs or Rb, allowing for experiments that reach quantum degeneracy for both species in parallel.

A comparison of the properties of the K D$_1$ line to other alkali-metal elements for which dRSC has not been implemented yet suggests that a transfer of our cooling scheme to these elements is promising. Especially the fermionic isotope $^{40}$K and the element Na have similar hyperfine splittings in the excited manifold. For $^{41}$K and Li with even smaller splittings the transfer might be more challenging. Recently, the combination of nondegenerate 3D Raman sideband cooling with high-resolution imaging systems \cite{Cheuk2015,Parsons2015} has allowed one to simultaneously cool and image individual atoms with single-lattice-site resolution. The cooling technique presented here can be extended to far off-resonant lattice configurations \cite{Vuletic98} and should allow simplifying the imaging techniques as presented in Refs.~\cite{Cheuk2015,Parsons2015}.

We are indebted to R. Grimm for generous support. We thank G. Anich, K. Jag-Lauber, F. Meinert, and G. Unnikrishnan for fruitful discussions. We gratefully acknowledge funding by the European Research Council (ERC) under Project No. 278417 and by the Austrian Science Foundation (FWF) under Project No. I1789-N20 (joint Austrian-French FWF-ANR project) and under Project No. P29602-N36.

\bibliographystyle{apsrev}
\bibliography{KCs-KRaman}

\end{document}